# Ultrasensitive Field-Effect Biosensors Enabled by the Unique Electronic Properties of Graphene

Xiaoyan Zhang,[1,2] Qiushi Jing,[1] Shen Ao,[1] Gregory Schneider,[2] Dmitry Kireev,[3] Zhengjun Zhang[1] and Wangyang Fu[1,*]

[1]*School of Materials Science and Engineering, Tsinghua University, Shaw technical science building, Haidian District, Beijing, 100084, P. R. China*
[2]*Leiden Institute of Chemistry, Leiden University, Einsteinweg 55, 2333CC Leiden, the Netherlands*
[3]*Department of Electrical and Computer Engineering, University of Texas at Austin, 78757, Austin, USA*

\* to whom correspondence should be addressed: fwy2018@mail.tsinghua.edu.cn

**Abstract**

This feature article provides a critical overview of current developments on nanoelectronic biochemical sensors based on graphene. Composed of a single layer of conjugated carbon atoms, graphene has outstanding high carrier mobility, low intrinsic electrical noise, however, a chemically inert surface. Surface functionalization is therefore crucial to unravel the graphene sensitivity and selectivity for the detection of targeted analytes. To achieve optimal performance of graphene transistors for biochemical sensing, in this particular review we highlight the tuning of the graphene surface properties via surface functionalization and passivation, and the tuning of its electrical operation by utilizing multifrequency ambipolar configuration and high frequency measurement scheme to overcome the Debye screening to achieve low noise and highly sensitive detection. Potential applications and prospective of ultrasensitive graphene electronic biochemical sensors ranging from environmental monitoring and food safety, healthcare and medical diagnosis, to life science research will be presented as well.

**Keywords**: graphene, biochemical sensors, field-effect transistors, ultrasensitive, electronic properties

# 1. Introduction

While the molecules always need to be specifically labeled for optical[1-3] or magnetic[4-7] based detection and analysis, label-free nanoelectronic biochemical sensing based on semiconducting techniques[8, 9] are more promising for portable point-of-care (POC) applications. Current research on such POC detection platforms has drawn worldwide interest, especially when driven by the concepts of Internet of Things (IoT), big data, and mobile health (mHealth). While versatile detection strategies exist, in order to fulfil all the requirements for a biosensor, the detection must be sensitive at clinically relevant concentrations of biomarkers as well as selective against various interferences that exist in biological samples.[10-12] In general, a specific chemical functionalization of the sensor surface with suitable biological recognition elements is required for selective detection. The sensitivity of a nanoelectronic biochemical sensor mainly depends on the immobilized receptor biopolymers, intrinsic electrical properties such as mobility and noise, as well as the sensing mechanism itself. Retrospectively, the so-called electric field effect is at the core technology of nanoelectronic biochemical sensing for detection of the charges that are introduced by a molecule. This field effect has been harvested to design the first generation of nanowire,[13] carbon nanotube,[14] and graphene-based field-effect transistor (GFET).[15] The experimental preparations and observations of the electric field effect in nanomaterials have inspired numerous experimental and theoretical works related to the application of nanoscale field-effect transistors (FETs) for high performance label-free biochemical sensors.[16-21]

Among various nanomaterials, graphene holds a special place due to its high sensitivity,[22, 23] i.e., the significant change in graphene's conductivity caused by charged biochemical molecules in direct contact with graphene. Due to a highly chemically stable planar $sp^2$ hybridization, the crystal lattice of graphene is intrinsically chemically inert. Therefore, graphene is not naturally endowed with stereospecific recognition of biomarkers. To achieve graphene biochemical sensors, various functionalization processes have to be utilized to introduce specific recognition moieties (e.g., antibody, antisense RNA, enzymes) onto the graphene surface to empower the recognition capability. The high sensitivity of graphene is attributed to its excellent electronic properties[15, 23] including room-temperature carrier mobility up to $10^6$ cm$^2$ V s$^{-1}$ and large specific surface area. Nevertheless, several scientific challenges have been recognized and actively pursued for field-effect type of biosensors based on charge detection. The challenges mainly root on the complexity of biochemical detection environments, i.e., the sensors are surrounded by the interferences of various external noise and have to overcome the Debye screening effect to achieve *in-situ* biosensing in physiological solutions; at the same time, the possible adsorption of background ions on the sensor surface and their effects on the sensing performance need to be investigated. Overcoming these challenges requires interdisciplinary research efforts in materials science, physics of semiconductor devices, chemistry, and biology. Along these lines, recent research trends on graphene nanoelectronic biochemical sensors now offer new opportunities for accurate measurements of human-based biomarkers at extremely low levels and/or monitoring trace amount of chemicals in environments,[19, 24] therefore providing great clinical value for early diagnosis and/or environmental monitoring and evaluation.

In this feature article we will cover the most recent developments in graphene-based biosensing devices, starting from basic principles of operation and moving towards current trends and future challenges. The discussion will begin with the advantageous electronic properties of graphene, i.e., the high carrier mobility, low intrinsic electrical noise, as well as the challenges ahead to achieve reliable biosensing operation of graphene electronic devices in biological environments. Functionalization of the chemically inert surface of graphene are at the core of graphene biosensors to unravel its sensitivity and selectivity performance for targeted analytes. In order to achieve optimal performance of GFET for biosensing, as illustrated in Figure 1, we split the approach in two strategies: one is tuning the graphene surface properties including surface *functionalization* and *passivation*; and the other is tuning the electrical *operation* of GFET. In this particular work we will highlight the utilization of multifrequency ambipolar detection and high frequency electrical field to overcome the Debye screening to achieve low noise and high sensitivity. Potential applications of GFET from environmental monitoring and food safety, healthcare, and disease detection, to life science research will be presented as well.

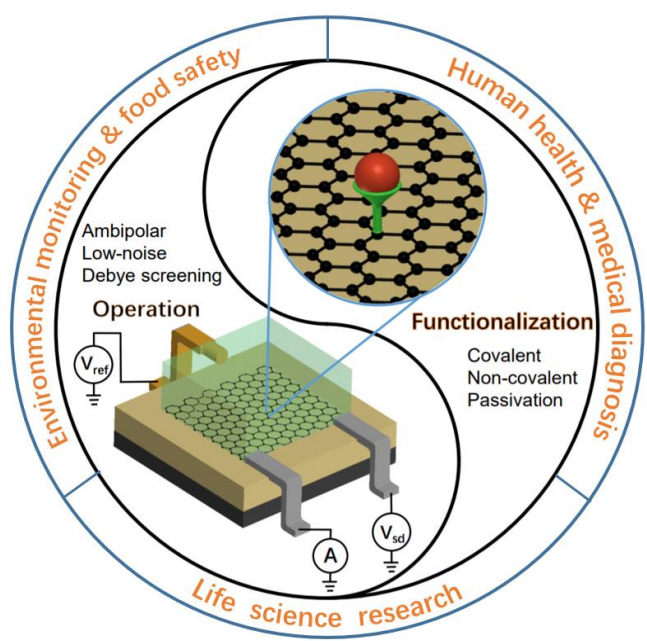

**Figure 1**. Operation, chemical functionalization and application of GFET devices that are both sensitive and selective.

## 2. Principle, fabrication, and operation of graphene nanoelectronics for biochemical sensing

Graphene is unique among solid-state materials as all carbon atoms are located on the surface and are extremely sensitive to environmental changes. Graphene electronic biochemical sensors have been explored to harvest not only the electronic properties of graphene for conductance detection in real time, but also its optical and mechanical properties for transparent and flexible sensor design,[25-27] its quantum capacitance for passive wireless sensing,[28-30] and even its low-frequency noise signatures with respect to various biomolecular adsorption for frequency domain detection.[31] Up to now, various sensing mechanisms such as charge transfer,[32] scattering,[33] capacitive effect,[28] and field effect,[18] have also been realized for highly sensitive detection based on graphene nanoelectronics. Particularly, in a semiconducting transistor device, field effect refers to the modulation of its surface conductance (or resistance) upon the application of a vertical electric field, and has been widely accepted as one of the most reliable sensing mechanisms. In this respect, we will focus on the recent progress of graphene field-effect biosensors. In Table 1 shows comprehensively the state-of-the-art performance of different type of graphene-based biochemical sensors. Nevertheless, we note here that the complexity of biochemical environment as well as the difficulty in understanding its interactions at the surface of biochemical sensors, make the origin of the sensor response not always fully clear, particularly disentangling the achieved sensing response from any kind of noise sources, for example nonspecific binding, surface chemical instability, drifts, to name a few. In addition, precautions have to be taken as surface condition (functionalization) of graphene as well as the device geometry might be different from sample to sample (and is not always reported in the literature). For example, annealing exfoliated graphene exhibited a comparable carrier mobility to as-fabricated exfoliated graphene (from ≈5000 $cm^2 V^{-1} s^{-1}$ to ≈5500 $cm^2 V^{-1} s^{-1}$), but also a strikingly decreased sensitivity to $NH_3$ gas (from ≈1 ppm to ≈1000 ppm) due to mainly the difference in their surface conditions. Therefore it might not always be fair to make such comparisons with different samples.[16, 34] For example, performing high quality biosensing (e.g., extracellular) measurements reproducibly require identical or nearly identical devices.[35, 36]

Table 1. Selected examples of the state-of-the-art performance of different type of graphene-based biochemical sensors.

| Type of graphene | Surface interaction (Surface modification & Target biomolecules) | Carrier mobility [$cm^2 V^{-1} s^{-1}$] | Sensing response | | Detection limit | Refs |
|---|---|---|---|---|---|---|
| | | | Concentration | Relative sensitivity ($\Delta R/R$ or $\Delta I/I$) | | |
| Exfoliated | 2D heterostructure (van der Waals) | ≈100000 | | | | [37] |
| | Bare graphene + BSA | | 300 pM | ≈2 % | 300 pM | [38] |
| | Iminobiotin(IB)-pyrene derivatives + avidin | 1000 | 1 pM | 60 mV($|\Delta V_{CNP}|$)* | | [39] |
| | Bare graphene + $NH_3$ | 5000 | 1 ppm | 300 % | 1 ppb | [16] |
| | Annealed graphene | 5500 | 1000~1800 ppm | <2 % | 1000 ppm | [34] |
| | ssDNA + chemical vapors | Dimethyl methylphosphonate | 1600 | 20 ppm | 2.5 % | ≈5.6 ppm | [40] |
| | | Propionic | | 90 ppm | 1.5 % | ≈31.3 ppm | |
| | | Methanol | | 7500 ppm | 1 % | | |
| | | Octanal | | 14 ppm | 1 % | | |
| | | Nonanal | | 0.6 ppm | 2 % | | |
| | | Decanal | | 1.7 ppm | 2 % | | |
| CVD | ssDNA + Complementary ssDNA | | | 10 nM | 11.2 % | ≈10 pM | [33] |
| | 20-mer DNA aptamer + ATP | | | 10 pM | ≈1 % | 10 pM | [41] |
| | HL-1 cells + action potentials | | 750 | | ≈1200 μV** | ≈200 μV | [36] |
| | Copper(I) + $C_2H_4$ | | 1500 | 1 ppm | ≈64 % | 2 ppb | [19] |
| | Penicillinase + penicillin | | | 3 mM | ≈0.6 % | 50 μM | [42] |
| | PEG and aptamer + IgE | | | 2.4 nM | 0.06 μA($\Delta I_{sd}$)** | 47 pM | [43] |
| | Anti-IL6(interleukin-6) + IL6 | | | 20 pg mL$^{-1}$ | ≈3.7 % | 2 pg mL$^{-1}$ | [44] |
| | Bare graphene +11-mer ssDNA | | 1200 | 100 pM | 3.0 % | 4 pM | [20] |
| | Pyrene-linked PNA + HIV DNA | | 1100 | 10 pM | ≈33 % | 2 pM | [24] |
| | Phenol + H$^+$ | | 2020 | pH=3 ~ 10 | 49 mV pH$^{-1}$($|\Delta V_{CNP}|$)* | ≈10$^{-5}$ pH | [18] |
| | Crown ether + K$^+$ | | 1520 | pK=0 ~ 4 | 60 mV($|\Delta V_{CNP}|$)* | | |
| rGO | PSA monoclonal antibody + PSA | | ≈30 | ≈1 nM | ≈ 17 % | ≈1.1 fM | [45] |
| | $Al_2O_3$ layer/Au NPs and antibody + Ebola glycoprotein | | | 1 ug mL$^{-1}$ | 17 % | | [46] |
| | OC antibodies + Amyloid-β fibrils | | | 1 pg mL$^{-1}$ ~ 10 ng mL$^{-1}$ | 1.8 % ~ 8.4 % | 1 pg mL$^{-1}$ | [47] |
| | Polyethylenimine/urease + urea | | 15~30 | 1~1000 μM | 9200 ± 500 μA cm$^{-2}$ per decade of [urea]** | 1 μM | [48] |

* Voltage shift of the CNP.
** Relative changes not given.

## 2.1. Principle of GFET for biochemical sensing

Graphene field-effect biosensors come from the big family of ion-sensitive FETs (ISFETs), which detect the conductance changes of the semiconducting channel upon binding of charged ions or biomolecules due to the field effect. To ensure a stable operation of the electronic sensor devices in electrolyte solutions, insulating layers such as $SiO_2$ and $Al_2O_3$ have been routinely adopted to isolate and protect the chemically reactive semiconducting channel from directly contacting the ions and biomolecules. Nevertheless, this relatively thick layer of insulating material also reduces the interfacial capacitive coupling between the sensor channel and the electrolyte solution, thus limiting the device sensitivity.

Since graphene is a conductor with every atom on the surface and is chemically inert, one may use it as the conducting channel in an ISFET and at the same time as the sensing surface to reach the highest sensitivity. Indeed, as a modern version of the classical ISFET, electrochemically gated GFET (and FET based on other two-dimensional (2D) materials) enables the detection of charged molecules in a label-free manner on a small footprint, and has demonstrated improved sensitivity compared to traditional bioassays.[16] An applied gate voltage either from a back gate or an electrolyte gate via a reference electrode or adsorption of charged molecules is able to shift the Fermi level ($E_F$) of the graphene layer, therefore modulating the conductance of the graphene device. Owing to the fact that graphene lacks an intrinsic band gap,[49, 50] typical transfer characteristics of the GFETs present an ambipolar behavior without an off-state. Generally, when sweeping the gate voltage $V_{GS}$ from negative to positive, the fermi energy $E_F$ of graphene shifts from the valence band to the conduction band. The resulting electron ($e$) and hole ($h$) conduction (respectively) leads to the ambipolar behavior. When $E_F$ crosses the Dirac point ($E_D$, or charge neutrality point, CNP), the type of charge carriers is altered and the graphene conductance reaches its minimum with a finite value. We define the transconductance as the derivative of source-drain current $I_{DS}$ with respect to $V_{GS}$: $g = \frac{dI_{DS}}{dV_{GS}}$, which reflects the resulting change in current in response to a small variation in the gate voltage, $V_{GS}$, and is one of the most important characteristics to evaluate the sensing response of a graphene transistor. Systematic studies on the GFETs fabricated on a variety of substrates with different width to length ratio $W/L$, reveal a linear relation between the source-drain current $I_{DS}$ (and thus the transconductance) and the $W/L$:

$$I_{DS} = \frac{W}{L} \cdot C_I \cdot \mu \cdot (V_{GS} - V_{CNP}) \qquad (1)$$

where $C_I$ is the interface capacitance, $\mu$ is the carrier mobility, and $V_{CNP}$ is the gate voltage at the CNP.

Conventionally, a GFET is favorably operated at its maximum transconductance $g_m$ to achieve the highest sensing response. As a result of the large interfacial capacitance and high carrier mobility, the value of the GFET transconductance can reach up to 200 μS,[35, 51] which is almost one order of magnitude larger than that of the other ISFET technologies based on Si or AlGaN materials. In practice, the maximum transconductance point of p-type doped graphene devices occurs in the hole conduction regime,[35] and vice versa for n-type doped devices. The doping effect could result from the water molecules trapped at the interface or from an unknown chemical doping,[35] induced by the external environment or process.[52]

Based on Eq. 1, to extract the value of the field-effect mobility, $\mu = \frac{L}{W} \cdot \frac{g}{C_I V_{DS}}$, we model the direct current (DC, or low frequency) interfacial capacitor $C_I$ of an electrolyte-gated GFET as two capacitors in series. One part is the intrinsic quantum capacitance of graphene, $C_Q$, which depends on the charge carrier concentration and can be determined as a function of the channel potential across the graphene sheet; and the other part is the virtual parallel-plate double layer capacitance, $C_{DL}$, formed due to the separation of the charges adsorbed on the graphene surface and the solution side of the interface as governed by the Poisson-Boltzmann equation (and independent to the gate voltage).[18, 53] In practice, $C_Q$ has its minimum value $C_{Q,min}$ and is directly related to the density of so-called effective charged impurities $n*$ (varying from $1\times10^{11}$ to $1\times10^{12}$ cm$^{-2}$), which represents the global behavior of defects and can cause local potential fluctuations in graphene.[53-55] Experimentally, we may obtain $C_I$ either by performing impedance measurements or capacitive Cyclic Voltammetry (CV) current measurements at different scan rates.

We note here that, for hydrophobic materials such as graphene, the air gap capacitance ($C_{airgap}$) should be included in the interface capacitance $C_I$.[56-58] Alternatively, a very recent research on the dielectric constant of water suggested the presence of an interfacial layer with vanishingly small polarization such that the out-of-plane dielectric constant of this very thin (~1.5-2 nm) confined water layer is only about 2.1 nm.[59] Interestingly, both hypothesises lead to an interfacial capacitance of about 1 μF cm$^{-2}$, which is in agreement with experimental results.

For liquid-gated GFETs, the major challenges lie in the control of the chemical functionalization, the identification of the exact sensing reactions at the graphene surface, and in the characterization of the number of charges each biomolecules carry. Usually, we assume a constant interfacial capacitance and carrier mobility $\mu$ of graphene upon biomolecular adsorption, which is correct in most cases where the targeted biomolecules adsorbed on the receptors and interact weakly with the underneath graphene lattice. However, additional scattering centers might be formed if biomolecules directly bind on a graphene surface. Such scattering centers are able to trigger a suppression of the mobility of charge carriers.[33] In addition, practical sensor designs should also take the possible changes of the interfacial capacitance upon biomolecule adsorption into account.[28]

**2.2. Fabrication and operation of GFET**

Usually, micromechanically exfoliated graphene has intrinsically higher quality with less defects.[34] Owing to the feasibility of large-scale fabrication at low cost, chemical vapor deposition (CVD) of graphene on metals thin films has been widely accepted as a more suitable technology platform for practical application compared to mechanical exfoliation. Additionally, further optimization of graphene biosensors also calls to study how the number of layers affects both the 1/$f$ noise and the sensing response to the surrounding environments, as the electrical properties of monolayer graphene are different from those of its few-layer counterparts[60]. In general, few-layer graphene devices feature less steep transfer curves (i.e., reduced transconductances) as compared to the single-layer graphene, resulting in substantial suppression in the GFET amplification. In this regard, monolayer graphene with a large sensing response is beneficiary.[23, 35, 61] The appropriate fabricating and packaging of graphene-based field-effect biosensors depends on their field of applications. In biotechnology, environmental monitoring, agriculture and food technology, the analysis of targeted analytes can be conducted in-line (e.g., in situ), on-line (e.g., discrete sampling) and off-line (e.g., in the laboratory). In general, the electrodes should be passivated or sealed with nonreactive materials to prevent any contact between the metal lines and the electrolyte, as well as to define the active gate area.[35] Technologically this requires an additional layer of a chemically stable resist, such as polyimide[62], polydimethylsiloxane (PDMS)[63] or solid state thin film such as $Al_2O_3$[64], $TiO_2$, $HfO_2$[65] and $Si_3N_4$[66] to avoid possible electrochemical processes between the electrodes and electrolytes, as well as any false signal due to the surface interaction with electrode materials. In medical and life science applications biosensors are generally categorized as *in vitro* and *in vivo*. For *in vivo* biosensors operated inside the body, the implants have to fulfil additional strict regulations on sterilization to avoid inflammatory response and on long-term biocompatibility to avoid harmful interaction with the body environment during the period of use.[64, 67] Based on the type of applied gate voltage, GFETs can be grouped into two major classes: so-called back-gated and liquid-gated GFET.

**2.2.1. Back-gated GFET**

A back-gated GFET consists source and drain metallic electrodes bridged by a graphene conduction channel (Figure 2a). To ensure a negligible contact resistance, electrodes (e.g., 5 nm Cr/50 nm Au) can be prepared directly on top of the graphene.[19] Double contacts are found to reduce the effective contact resistance. Usually, GFET devices are fabricated by transfer of CVD graphene onto highly doped conductive silicon substrates with silicon dioxide insulating layers. The carrier density, and the corresponding channel conductivity can be modulated by applying potential to the highly conductive silicon substrate to a range of back-gate voltages $V_{GS}$ via field effect. In a typical measurement, one applies a constant source-drain bias voltage, $V_{DS}$, and monitor the resulting source-drain current $I_{DS}$ between the source and the drain of the graphene channel, when changing the back-gate voltage $V_{GS}$. However, GFET devices fabricated on bare $SiO_2$/Si substrate are almost always haunted by intensive $p^+$-doping (i.e., CNP shifts to more positive voltages) and large hysteresis due to the trap states at the graphene/$SiO_2$ surface.[68] One possible way to avoid this unwanted doping is to transfer graphene onto surfaces treated with hexamethyldisilazane (HMDS)[69] or octadecyltrichlorosilane (OTS) to shield it from trapped charges located at the $SiO_2$ surface, resulting in advantageously hysteresis-free operation and close-to-zero CNP point. Such high-performance back-gated GFETs with reliable operation have been successfully used for gas sensors.[19, 23] When the back-gate is held at a fixed voltage, physisorption or chemisorption of targeted molecules on the graphene surface can induce a change in the electric field, and thus the channel current due to field effect.

**2.2.2. Liquid-gated GFET**

In comparison to the back-gate geometry, where the gate voltage is applied to the highly conductive silicon substrate, in a liquid-gate configuration a reference electrode together with the electrolyte serves as the 'gate electrode' (Figure 2b). The liquid gate is coupled to the graphene channel through the interfacial capacitance $C_I$ as introduced previously.[70]

During the fabrication of a GFET for liquid-gate operation, a passivation is required in order to prevent any contact between the metal lines and the electrolyte as well as to define the active gate area.[35] Technologically this means additional layer of chemically stable resist, such as polyimide or epoxy.[36] In a typical example, a biocompatible, two-component epoxy were applied after wire bonding for sealing the metallic electrodes against the liquid-gate voltage $V_{ref}$ applied via the Ag/AgCl reference electrode, to prevent any possible leakage current in electrolyte environment.[36] We note here that owing to the unique frequency dependent dielectric properties of the water solution, electrolyte gating is capable of tuning the conductance of GFETs without shunting the propagating radio frequency (RF) signal.[71] For life science application, a new passivation type has been recently introduced and argued to provide a better interface between the GFETs and neuronal cells.[36] This 'feedline follower' passivation covers only the area over the metallic feedlines, thus helping to prevent membrane bending as the neuron cell approaches the graphene and grows consistently on it.

For comparison, the typical area normalized transconductance of an electrolyte-gated GFET is reported in the range of 1-2 mS V$^{-1}$.[36] Figure 2c-e depict a typical GFET biosensing measurement. In Figure 2d, a receptor molecule is immobilized on the surface of graphene for specific recognition of target biomolecules. The binding of a negatively charged molecule causes a positive shift of the $I_{DS}(V_{ref})$ curve due to the field effect. In the time-dependent measurement as shown in Figure 2c, such positive shift of the transfer characteristics of the GFET results in an increase of the current $I_{DS}$ in the hole regime (as indicated by 'h' in red), and a decrease of the current in the electron regime (as indicated by 'e' in light blue), and vice versa for the detection of positively charged molecule (Figure 2d). In this respect, in principle no sensing response of GFET sensors is expected upon the binding of non-charged biomolecules, unless a charge variation can be introduced through subtle dipole fluctuation[72] or molecular engineering[73].

Usually, even after the treatment of substrates (e.g., the $SiO_2$ treated with HMDS[69] prior to graphene transfer to support the graphene from trapped charges on the surface of $SiO_2$), it is common that multiple neutrality points are observed together with relatively large hysteresis if measured against liquid-gate voltage sweeping.[24] These poor device performances indicate the presence of a large amount of surface contaminants or charged trap states at the graphene/electrolyte interface, even though ~200°C baking and thoroughly rinsing in isopropanol are routinely applied. To overcome this issue, an *in-situ* electrochemical cleaning method can be adapted for graphene surface refreshment.[24] Every consecutive cleaning cycle removes the spurious neutrality points and decreases the hysteresis, and usually after 10 cycles of refreshment, the $G(V_{ref})$ curve of the GFET becomes completely stable, and both the spurious neutrality points and the initial hysteresis observed can be eliminated (Figure 2f).

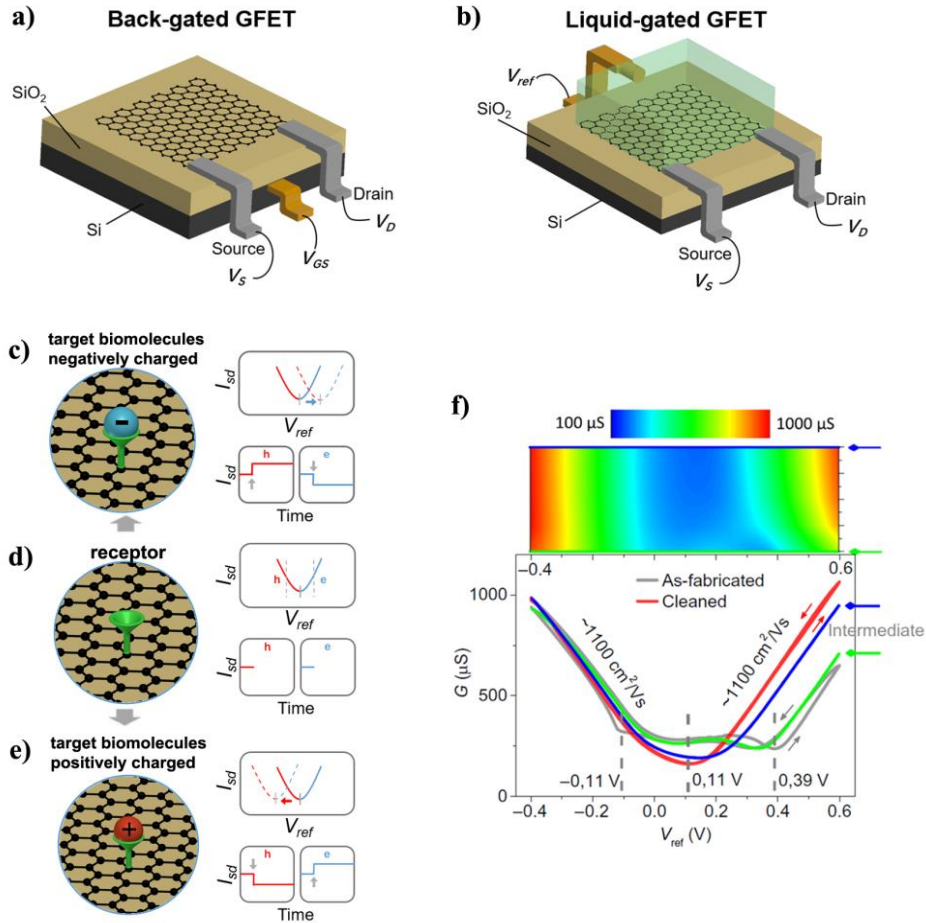

**Figure 2**. Schematic representation of (a) a back-gated GFET and (b) a liquid-gated GFET. (c-e) show the sensing principle of the liquid-gated GFET biosensor. (f) Upper panel shows the plot of a GFET sheet conductance after electrochemical cycles under an operation of $V_{ref}$ between -0.4 V and 0.6 V. Lower panel shows the $G(V_{ref})$ curves before electrochemical cycle (line in gray), during the first electrochemical cycle (line and arrow in green), and after 5 times and 10 times electrochemical cycle (line and arrow in blue and red), respectively. Both hole and electron carriers represent symmetric ambipoar behavior with carrier mobility of ~1100 $cm^2/V_s$.

### 2.2.3. Bandgap engineering

The lack of electronic band gap of graphene leading to the metallic nature of graphene-based FET devices, that cannot be switched off at room temperature. In order to overcome the limitation caused by the zero band gap structure, the engineering of bandgap is one of the most critical points for graphene-based digital devices. In this respect, graphene derivatives, including graphene nanoribbons (GNRs), heteroatoms (N, S, B, P) doped, functionalized, and bilayer graphene, were explored to modulate the electronic structure of graphene and improve the on/off current ratios of GFETs. Heteroatom-doping is a process that some carbon atoms in the graphene structure are replaced by the heteroatoms. The size and electronegativity of the heteroatoms are often different from those of carbon atom. Therefore, regardless of whether the dopants have a higher (as N) or lower (as B, P, S) electronegativity than that of carbon, the introduction of heteroatoms into graphene carbon networks could cause electron modulation on the charge distribution and electronic properties of carbon skeletons, which in turn affects their performance for electronic applications.[74] Besides chemical modification, graphene nanoribbon, graphene nanomesh, and graphene nanoring,[75] have also been proved as rational designs of the graphene to open a bandgap, yielding an improved transistor $I_{on}/I_{off}$ ratio. For example, GFETs prepared with sub-10-nm GNRs showed high on-off current ratios of $10^4$.[76] By using a rapid-heating plasma CVD, which is accessible for large-scale production, high-yield GNRs (250 000 $cm^{-2}$) were prepared.[77] Remarkably, by the on-surface bottom-up approach, atomic level defined GNRs were synthesized with designed halogenated aromatic precursors. Depending on the structure of precursors, GNRs with defined width and edge-type can be achieved by converting monomers through dehalogenation and coupling reactions.[78, 79] Recently, tuning the band-structure of graphene superlattices with hydrostatic pressure,[80] and even achievement of unconventional superconductivity and insulator behaviour with magic-angle[81, 82] were realized. The diversified approaches would promote the development of graphene-based electronic performance in logical circuits. Nevertheless, notably, the transistor $I_{on}/I_{off}$ ratio has no direct relation to the performances of a sensor device, although it is related to graphene digital applications requiring high on state current ($I_{on}$) and ultra-low power consumption at the off state ($I_{off}$) of the transistors.

**2.3. Challenges in GFET for achieving ultimate single molecular detection**

Granted by the excellent electrical properties,[83, 84] including extraordinary high mobility[22, 85, 86] and low intrinsic electrical noise,[87, 88] which gives a better signal-to-noise ratio (SNR),[56, 89-91] graphene-based biochemical sensors are reported to provide superior performances compared to their Si-based counterparts and/or traditional bioassays.[61] However, there are challenges that remain to be solved with systematic and comprehensive research. Current reported GFETs still cannot achieve the theoretical predicted performance, which can be related to several basic and important properties of graphene affect the performance (especially electrical noise) of GFET. In the following section, starting from general considerations regarding the carrier mobility of graphene, we will focus on the electrical noise and Debye screening that hinder the practical application of GFET for ultrasensitive detection under high ionic or physiological solutions.

**2.3.1. Carrier mobility**

The sensing response of GFET is defined as $S = \Delta I_{DS}/N$, that's a minute field effect (the electrical current $\Delta I_{DS}$) induced by the binding of a biomolecule carrying electron charge $Ne$. According to Eq. 1, $S$ is therefore proportional to the carrier mobility $\mu$ of graphene, as well as the slopes of the sublinear $I_{DS}(V_{GS})$ curves (i.e., transconductance $g$). In this respect, if the values of other parameters including the electrical noise amplitude are equal, a higher carrier mobility $\mu$ (up to $10^6$ cm$^2$ V$^{-1}$ s$^{-1}$ under room temperature) implies a better sensor performance upon the adsorption of charged biomolecules. Although there are no direct evidences or theories that could unambiguously relate the high mobility of GFETs to their noise performances, a higher carrier mobility indeed complies with graphene bearing less impurities and defects, and therefore being favor of an improved noise performance. Applications that could exploit these unique properties such as label-free electronic biochemical sensors with ultrahigh sensitivities will be introduced in Section 5. Specifically, owing to its exceptional high mobility, graphene is potentially well suited to RF applications,[71] and holds great promise for sensing applications at a high sampling rate where a wide bandwidth is of key importance.[18]

Since the sensing response of GFET sensors depends on the carrier mobility $\mu$, it is preferential to use and integrate high quality graphene into devices. The factors that affect the carrier mobility of graphene are listed as follows. *i*) Among various synthesis methods, micromechanical cleavage yields graphene with less defects, and therefore higher carrier mobility and lower intrinsic electrical noise.[92] Generally, carrier mobility in the order of $\approx$3 000-15 000 $cm^2\ V^{-1}\ s^{-1}$ are routinely reported for exfoliated graphene on $SiO_2$/Si wafers,[93, 94] in comparison to $\approx$100-1500 $cm^2\ V^{-1}\ s^{-1}$ of silicon materials.[95] The CVD grown graphene-based FETs would feature mobility in the order of 1 000-10 000 $cm^2\ V^{-1}\ s^{-1}$. However, $\mu$ of CVD grown graphene can be substantially improved (50 000-350 000 $cm^2\ V^{-1}\ s^{-1}$) by using single-crystal graphene free of grain boundaries[96, 97] transferred onto a high-quality hexagonal boron nitride (*h*-BN) substrate.[98-100] Such high $\mu$ of CVD graphene even rivals those of exfoliated samples, making the CVD method ideal for the synthesis of large-area, high-quality graphene for sensing applications. *ii*) The influence of surface functionalization on the electronic performance of the GFETs has been studied. To a large extend, the exceptional electrical properties of graphene can be preserved during the noncovalent chemical treatment process. For example, the transfer characteristics $G(V_{GS})$ of these GFET devices exhibit symmetric shapes and field-effect hole carrier mobilities of ∼1500 $cm^2\ V^{-1}\ s^{-1}$, which are preserved to approximately 80% of their initial values upon the noncovalent surface functionalization with the copper(I) complexes via π-π and/or hydrophobic interactions. The affordable drop in the carrier mobility was ascribed to an increased scattering of the charge carriers.[19] As it was experimentally proven recently, noncovalent functionalization can indeed deliver GFETs with fully preserved mobility.[18] By using aromatic noncovalent functionalization,[18] the hole and electron mobilities of this phenol-activated GFET for pH sensing were found to be 1770 $cm^2\ V^{-1}\ s^{-1}$ and 2020 $cm^2\ V^{-1}\ s^{-1}$, respectively. These mobilities, as well as those of the fluorobenzene-passivated GFET (2650 $cm^2\ V^{-1}\ s^{-1}$ for hole and 3260 $cm^2\ V^{-1}\ s^{-1}$ for electron), show no degradation and are order-of-magnitude higher than those of high-performance ISFETs formed on silicon-on-insulator (SOI) wafers.[18] On the other hand, covalent surface functionalization, for example, the addition of only one H-sp$^3$ defect per ~250,000 down to ~145,000 sp$^2$ hybridized carbon atoms (correspondingly decreasing carrier diffusion length, $L_D$ from 45 nm down to 35 nm) effectively affects the mobility of charge carriers in graphene compared to pristine graphene. Nonetheless, for sensing applications, the reduced carrier mobility of highly hydrogenated graphene is still sufficient.[101] *iii*) In fact, the extremely high mobility values only happen in devices with small channel areas (below 100 μm$^2$), which may due to a finite crystallinity of the graphene. In smaller devices there are fewer grain boundaries and defects, reaching a situation when a GFET consists of a single graphene crystal. In this case, a drastic increase in that transistor's charge carrier mobility is expected. While for the GFETs with channels over 100 μm$^2$ in area, the chance of meeting grain boundaries and defects increases, restricting the electrical performance of the GFETs,[102, 103] although some researchers have specifically used the grain boundaries for e.g., ion channel sensing applications.[104] *iv*) Scattering induced by substrates that constricts the electrical properties of even single-crystalline graphene, is another affect for the observed limited mobility values. For example, the SiO$_2$ substrate results in a suppression of the average mobility 750 ± 350 cm$^2$ V$^{−1}$ s$^{−1}$[105] compared to HfO$_2$ and polyimide substrates, whose value reaches up to 4.9 × 10$^3$ cm$^2$ V$^{−1}$ s$^{−1}$. Carrier mobility up to 10 000-197 600 cm$^2$ V$^{−1}$ s$^{−1}$ was achieved by encapsulating graphene in *h*-BN,[37, 106, 107] providing unprecedented possibilities for sensing applications if considering recent progress towards the growth of large-area high-quality, single-crystal graphene[108] and *h*-BN monolayer on Cu.[109]

**2.3.2. 1/*f* noise**

GFETs have exceptionally high carrier mobility, which results in high transconductance that endows the sensors with a significant current response to minute changes in the surface potential of graphene caused by the adsorption of molecules. However, to determine the maximum sensitivity of GFET, it is essential to investigate the electronic noise performance, which is ubiquitous in solid-state electronic devices and sets a limit on the smallest signal that can be possibly resolved. Generally, the inherent 1/$f$ noise dominates the electronic noise of GFET at biologically relevant low frequencies (≲1000 Hz). Such low-frequency 1/$f$ noise, whose power spectral density (PSD) inversely depends on the frequency $f$, is governed by surface over bulk noise in graphene up to seven layers[60]. As valuable tools for predicting the detection limit of biochemical FET sensors, noise measurement and characterization are well-established in the MOSFET community. It is revealed that the 1/$f$ noise in graphene largely depends on the number of layers. For monolayer graphene supported on a $SiO_2$/Si substrate, its 1/$f$ noise is comparable to that of bulk semiconductors (including Si).[110] Double- or few-layer graphene devices are expected to reduce the 1/$f$ noise because the potential fluctuations from external charged impurities such as oxide traps and/or interface states can be effectively screened.[60, 110] By comparing the noise performances of a $SiO_2$/Si substrate supported GFET device and its counterpart after etching the underlying $SiO_2$ substrate to suspend the monolayer graphene, one order of magnitude reduction on 1/$f$ noise was observed.[88] Since the 1/$f$ noise in monolayer graphene is a surface phenomenon, such dramatic reduction is mainly attributed to the removal of the supported $SiO_2$ substrate, and thus any accompanied external trap states.[60]

Defects in the graphene lattice is another origin of noise. For example, compared to scotch-tape exfoliated graphene or CVD grown graphene, the permanent oxygen-based defects are introduced by over-oxidation of graphene oxide (GO). An incomplete removal of oxygen groups for reduced graphene oxide (rGO) also leads to degradation in the mobility and noise performance.[111] Interestingly, the large concentration of defects of GO (and rGO) leads to improved sensing responses when used as an active sensing electrode, compared to near defect-free exfoliated monolayer graphene.[111, 112] In principle, optimal defect density can be achieved by balancing the gains in the sensing response against the rapidly degraded low-frequency 1/$f$ noise when increasing the density of defects.[111] In reality, the challenges lie in controlling the density of the defect, particularly when lacking of knowledge on the nature of the defect. Remarkably, either environmental exposure or aging of graphene devices increases the level of noise; in contrast, a proper capping layer or surface functionalization circumvents can even reduce the level of noise.[113] For example, by encapsulating a monolayer graphene between two sheets of *h*-BN, the channel area normalized PSD can be suppressed up to one order of magnitude lower compared to its non-encapsulated counterparts.[114]

The sensitivity limit of liquid-gated GFETs can also be indicated by an root-mean-square (RMS) value of the gate voltage, which can be directly derived from the measured noise PSD and the device transconductance.[35] Liquid-gated GFETs should be capable of detecting single voltage spikes caused by a cardiomyocyte as low as 100 µV with a SNR above 10, which outperform FETs made from conventional materials such as Si and is comparable with that of the state-of-the-art recording systems based on microelectrode arrays (MEAs).[35]

The sources of noise in the liquid-gated GFETs are still not fully understood. Nevertheless, the current noise PSD shows a minimum at the CNP, which increases when moving away from the CNP for $|V_{GS}| \gg V_{CNP}$.[115, 116] Further systematical investigations on the 1/$f$ noise behavior of graphene devices fabricated on different substrates, suggested a 'V'- or 'M'-shaped feature of the noise amplitude regardless of the substrates (SiO$_2$, Si$_3$N$_4$, and sapphire).

As a special application, the low-frequency electronic noise can be advantageously adopted for realizing selective graphene gas sensors.[31, 117] The mechanism is based on the distinct noise features of the graphene transistors upon the adsorption of various vapors of different chemicals. This sensing mechanism achieves selective frequency domain detection without specific surface functionalization of graphene, and calls for future exploration for other 2D materials.

**2.3.3. Electrochemistry**

In liquid-gated GFET biosensors, the electrical current should be confined transversely in the graphene conductive channel, which can sustain a high current density over 1000 µA µm$^{-1}$ before Joule-heating breakdown.[118] Whereas spurious electrochemical current due to redox reaction at the graphene/liquid interface flows vertically and interferences the sensing performance of gate controlled GFET devices.[17] To suppress electrochemical processes and the resulting exchange ionic currents, generally GFETs are operated at low electrolyte-gate voltage, where the interface is considered to be inert and purely capacitive. Although it is not always explicitly stated in most of the literature, the design of electrochemically-gated GFET for detecting targeted analytes in real time greatly depends on our ability to understand and maintain a low level of electrochemical current. Basically, the electrochemical current (or gate leakage current) $I_{GS}$ increases with the gate source voltage $V_{GS}$. For example, application of high $V_{GS}$ potentials up to 1.4 V via an Ag/AgCl electrode, increases the gate leakage current up to tens of nA. Such experimental artifacts at moderate or relatively high electrolyte-gate voltages are considered of electrochemical nature, rooting on the exchange ionic current between the graphene channel and possible redox active molecules in the solution phase. As we will discuss in details in Section 3.2, where we will probe into the interplay between the electrical in-plane transport and the electrochemical activity of graphene, it is possible to maintain a lower level of the gate leakage by tuning the density of H-sp$^3$ defects introduced by using plasma treatment.[52]

On the other hand, as we introduced previously in Figure 2f, an electrochemical gate leakage current can induce consecutive cleaning and improving the performance of the GFETs.[36]

**2.3.4. Debye Screening**

Debye screening effect describes the tendency of plasma to eliminate internal electrostatic field. As a large system containing mobile ionic charges, an electrolyte solution can be regarded as plasma. The screening layer is composed of diffusive movable ions attracted to a charged surface via the Coulomb electrostatic force. At room temperature, the Debye length[119] can be formulated as: $\lambda_D = 0.304/I^{1/2}$, where $I$ is the ionic strength, and is typically ≈0.7 nm under physiological conditions. Given the size of biomolecules is in the range of several nanometers, it is therefore unlikely that they can approach the sensor surface within the Debye length to be recorded by the transistor. For example, for biotin receptors anchored on the nanowire surface with near side distance of ≈1 nm, no response can be observed upon the binding of streptavidin from a 10 nM solution under physiological conditions (1×PBS, $\lambda_D$ ≈ 0.7 nm).[119] At low salt concentrations of 0.1×PBS, the hybridization of complementary DNA molecules exhibits a normalized resistance change of 80%. When increasing the buffer concentration to 1×PBS, such sensing response decreases dramatically to 12%, which was found to follow the Debye length considerations.[120] A full screening of the biological binding signal happens when increasing further the buffer concentration to 10×PBS, resulting in negligible sensing response even at a relatively high complementary DNA concentration of 1 μM. So in fact, the Debye screening effect has put a fundamental obstacle to the possible sensing applications of the GFETs (and ISFETs in general) for real-time detection of relatively large biomolecules at high-salt/physiological conditions, although in principle GFETs are sensitive to changes below one single charge.[16, 121] Indeed, various approaches have been pursued to circumvent the Debye screening effect towards ultimate detection of biomolecules, including rational design of short antibodies, *ex-situ* measurement in low ionic strength buffers, and incorporation of porous polymer layers permeable to biomolecules.[41, 119, 122-124] Nevertheless, to achieve highly sensitive, real-time detection in high-salt/physiological conditions without any specific aptamer molecular design and restriction on interface environment, more general and novel operation strategy still have to be developed.

As a new sensing technology, radio frequency (RF) method is not restricted by Debye screening, and can detect biological molecules deep into the solution. This is because at RF frequencies (starting from ~10 MHz), due to the viscous effect of the solution, the movement of mobile ions in the electrolyte will lag behind the change of the AC electric field.[70] Therefore, aqueous solutions can be considered as dielectrics at high frequencies, and the Debye screening can be overcome. As a result, the binding events of biological molecules on the sensor surface can be detected as the changes of capacitance (or dielectric environment) caused by the replacement of water molecules with very high dielectric constant (= 80) by the target biological molecules with low dielectric constant (= 3-5). The development of RF sensors for biological detection is the current research trend. So far, most RF sensing schemes for biomolecular recognition, including biotin-streptavidin, DNA hybridization and glucose, have been demonstrated based on RF passive devices and circuits (e.g., various resonators and interdigital circuits), with a focus on basic research to test and study the interaction between RF and biomolecules to further improve the performance of the biosensor.[125] For example, the University of California, Davis, reported an RF resonant sensor based on a metal transmission line, which can detect the binding and activity of biological molecules at depths ranging from 10 nm to the surface of the sensor in a physiological solution environment.[126] The University of Michigan reported that carbon nanotube-based mixers could detect streptavidin macromolecules.[72] Twente University used high-frequency CMOS capacitance detection technology to achieve imaging of living cell networks at the single or even subcellular level.[127] However, it should be noted here that high frequency signals can deeply diffuse into buffer solution, which makes this sensing scheme more vulnerable to external environmental noise and RF detection of single biomolecule has not been realized.

**3. Graphene surface property tuning in graphene biochemical sensing**

Ideally, the crystal lattice of graphene is free of dangling bonds and intrinsically chemically inert. Therefore, to achieve graphene biochemical sensors, specific recognition moieties (antibody, antisense RNA enzymes, etc.) have to be introduced via both covalent[128-130] and noncovalent[131-133] approaches. Chemical functionalization of graphene surface using different biochemical molecules and chemical treatments, not only is essential for unlocking its sensing potential, but also plays a vital role in surface passivation of graphene against unintended non-specific binding to achieve high sensitivity and selectivity in high ionic background levels (Figure 3), which will be discussed in the following sections.

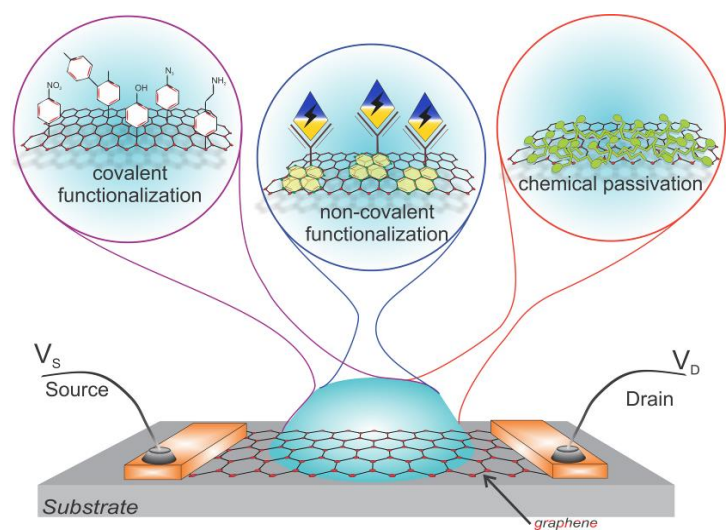

**Figure 3.** An illustration of graphene surface property tuning for biochemical sensing.

## 3.1. Functionalization of graphene for biochemical sensing applications

### 3.1.1. Covalent functionalization

Chemical functionalization of graphene is routinely achieved using either covalent[111, 115, 116, 128-130, 134] or noncovalent[40, 131-133] strategies. Covalent chemical modification[135] reliably modify the graphene surface by reacting with the $sp^2$ carbon atoms in the aromatic lattice. The covalent approaches allow engineering the properties of graphene with respect to its band gap and biointerfacing to a large extend. However, the resulted $sp^3$ centers at the reaction sites jeopardize the aromaticity of the graphene lattice and yield inferior electrical mobility and noise performance compared to pristine graphene. Covalent functionalization process also reveals the possibility to continuously transform graphene – a highly conductive zero-band gap semimetal – into an insulator known as graphane[128] or 2D Teflon.[129, 136] The reaction efficiency depends on parameters including the number of graphene layers,[137] the electrostatic charges,[138] and the defect density.[139] GO (or rGO) is a typical example of graphene materials resulted from covalent modification of the graphene scaffold with oxygen functional groups (e.g., carboxyl, hydroxyl and epoxy moieties) by using oxidative reactions (or with a chemical reduction step for rGO).[140] As a result of the large concentration of defects in comparison to near defect-free exfoliated graphene, GO and rGO show improved sensing responses yet inferior field-effect properties.[111, 112] Particularly, without damaging the lattice integrity and the resilient basal plane, halogenated graphene[141] (include hydrogenated graphene and fluorinated graphene, etc.) are promising for progressively tweaking graphene with $sp^3$ defects by introducing atomic hydrogen or fluorine into the honeycomb graphene scaffold. Regarding sensing applications, calculations predicted that partially hydrogenated graphene has a high affinity for $NO_2$,[142] while fluorographene can be applied for the detection of ammonia,[143] ascorbic acid, and uric acid.[144] The fluorine-enriched material, on the other hand, could also be adopted for genosensing upon further functionalized with thiol groups.[145] Cyanographene and graphene acid represent newly developed graphene derivatives,[146] which are promising for electrochemical sensing for the detection of biomarkers (e.g., ascorbic). Particularly, positively charged cyanographene exhibits a higher affinity for negatively charged analytes due to the electrostatic attraction compared to negatively charged graphene acid. In this respect, the optimisation of graphene derivatives (in the nature and the concentration of the functional groups, etc.), is mandatory for achieving sensing applications with high selectivity and sensitivity. Besides the defects on the basal plane of graphene, the edge of graphene represents another type of defects, which also plays an important role in the determination of its electronic and chemical properties for sensing applications. For example, defective rGO can be achieved by using enzymatic oxidation followed by hydrazine reduction. Such prepared rGO samples contain abundant edge defects and exhibit a prominent and selective sensing response towards the detection of hydrogen, particularly when activated with catalytic Pt nanoparticles.[147]

**3.1.2. Noncovalent functionalization**

Alternatively, noncovalent functionalization has the major advantage of fully preserving the aromatic lattice and thus the electrical performance of graphene lattice,[18, 19, 23, 135] for applications including band gap engineering, controllable n- or p-doping of GFETs, and linker molecular design. Noncovalent bond achieved via aromatic molecules can also be quite strong. For instance, the π-π interactions of graphene-benzene leads to a considerable binding energy of about 0.1 eV per carbon atom, and the binding energy of graphene-tetraphenylporphyrin can be estimated as 3.2 eV, i.e., approximately 90% of the typical binding energy of covalent C-C bond (≈3.6 eV).[148, 149] However, compared to covalent functionalization, noncovalent functionalization is believed to be less compatible with long term usage, where the stability and reliability are of key importance. Nevertheless, the weaker interactions of noncovalent functionalization could also be an asset for regenerating and recycling the sensor surface and thus the sensor devices.

Generally speaking, noncovalent approaches can be classified based on their intermolecular interactions with graphene, including π-π or hydrophobic interaction, electrostatic interaction, and van der Waals stacking.[135] The corresponding molecular self-assembly process on the surface of graphene could be accurately controlled in favor of an actual sensor design.[150] Functional molecules with a specific aromatic linker group (e.g., a pyrene unit) can be anchored onto graphene surface noncovalently via π-π and/or hydrophobic interaction, which is robust upon exposure to ambient conditions.[151] For example, a synthetic complementary peptide nucleic acid (PNA) molecules (5′-AAG CTA CTG GA-Lys (Pyrene)-3′) with a pyrene group has been introduced onto a graphene surface, which can perform as a receptor to target HIV virus related 11-mer ssDNA molecules (5′-TCC AGT AGC TT-3′, a fragment of the HIV-1 Nef gene29).[20] The final graphene surface features firmly adsorbed ssDNA molecules via π-π interaction.[40, 61, 152] Such a noncovalent π-π stacking between the pyrene linker group of the ssDNA molecules and graphene sheet results in an increase of the CNP resistance from 23.6 kΩ to 24.3 kΩ, suggesting a slightly increased carrier scattering due to pyrene-linked ssDNA molecules in the vicinity of the graphene. Also, chemical gating[61] by the negative charges of the ssDNA molecules causes a positive shift of the transfer curve (+150 mV),[20] corresponding to an adsorbed total negative charge density of −2e per 100 square nanometers.[56] Nevertheless, most of the previous studies have not taken into consideration, for example the effects from the conformational change of attached/bounded molecules, which may significantly influence the sensing response of GFETs. As an example, study on the RNA hairpins on a graphene surface indicated that biomacromolecules can display very different behaviours depending on the surface hydrophobicity, concentration of RNA and temperature.[153] Besides DNA, proteins[154-157] or peptides[131, 132] containing aromatic moieties could also self-assemble on a graphene scaffold.[132] Besides the charges that the biomolecules carry, depending on the Hammett constants $\sigma_p$, charge transfer from these molecules to graphene would result in a shift of the Dirac point before and after the surface functionalization of graphene.

Phenyl rings can also be included in the ligands functionalized with trifluoromethyl groups. Such aromatic moiety helps to stabilize copper complex in its monocationic state for ethene sensing, as well as to induce π-π stacking interactions to enhance the electronic coupling and thus the attachment between the sensitizer molecules and graphene.[19] Although the field-effect mobility is reduced to approximately 80% of the initial values upon the self-assembly of the copper(I) molecules on the surface of graphene (as we discussed previously in Section 2.3.1), no trend in the change of the Dirac points before and after the functionalization has been found. Therefore, it is likely that the differences in scattering rates caused by the organization of the complexes on the surface of graphene obscure the more subtle field effect of the complexes of different polarities.[19]

In these studies, noncovalent functionalization has demonstrated its suitability as nondestructive process for engineering the property of graphene. However, to what extend the outstanding electrical properties of graphene can be preserved, as well as what detection limit can be achieved by using such graphene sensors, are not always clear. To answer these questions, researchers configured GFETs with aromatic molecules containing hydroxyl (OH) groups (phenol), which protonate or deprotonate when decreasing or increasing the pH values of the buffer solutions, respectively. Such GFET sensors demonstrated fully preserved high mobility and exhibited a significant pH response (see also Section 2.3.1). Therefore these results increase the credibility and fidelity of high-performance graphene biochemical sensors and expand their applications for potential development of ultrafast detection at a high sampling rate where a high mobility is of key importance.[18]

Additionally, we can exploit the weak van der Waals-like interaction between layers to sandwich graphene with other 2D layers, including $MoS_2$, mica, or $h$-BN, to adjust and to achieve astonishing electronic properties. Such process is also called 'encapsulation'. For example, encapsulating graphene in a h-BN stacking layer can achieve high carrier mobility up to 140 000 $cm^2$ $V^{-1}$ $s^{-1}$, which is close to the theoretical limit imposed by acoustic phonon scattering at room temperature. Such extraordinary high value can be ascribed to very clean interfaces between graphene and $h$-BN, as well as their perfect lattice matching and effective screening of all the defects and roughness.[106] Similarly, the tunnel barrier for graphene spintronics can also be realized by CVD h-BN, placed over the graphene and used either as a monolayer or bilayer.[158] Moreover, graphene-$MoS_2$-metal hybrid structures can be used as ultrasensitive plasmonic biosensor.[159] MoS$_2$ as well as other 2D materials by themselves provide additional possibility for noncovalent functionalization as routes towards novel field-effect based biosensors.[160-162]

**3.2. Passivation of graphene surface to achieve high selectivity**

As previously discussed, surface functionalization including noncovalent and covalent approaches are essential to unlock the sensing potential of graphene surface, but it is important to keep in mind that chemical passivation plays a critical role for surface functionalization of graphene to realize high selectivity.

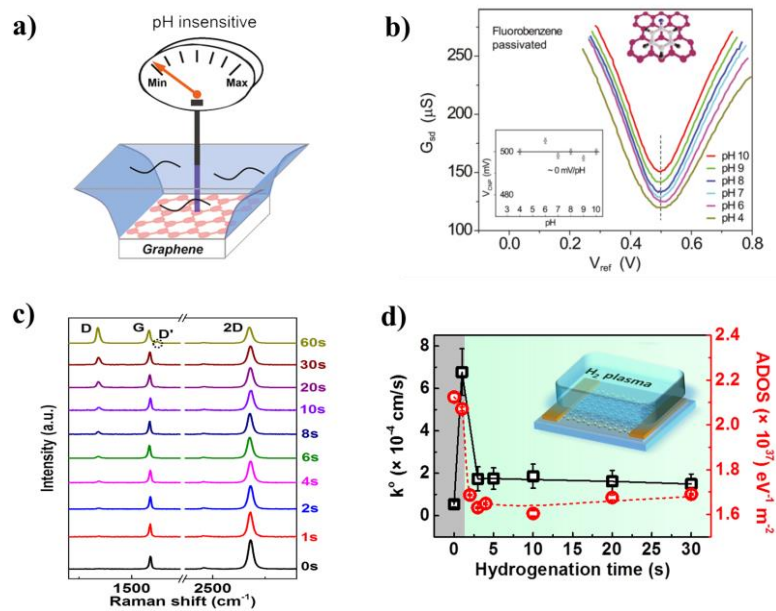

**Figure 4**. (a) GFETs are inert to pH variations in solutions. (b) $G_{sd}$ versus $V_{ref}$ of a GFET functionalized with fluorobenzene and measured in buffer solutions with different pH values. Inset indicates the CNP of the transfer curves keeps constant under different pH. (c) Monolayer graphene Raman spectroscopies under hydrogen radical plasma treatment of different duration. (d) The electron transfer rate $k^0$ and average DOS (ADOS) of graphene under hydrogen radical plasma treatment with different time.

Here we specially focus on the passivation of active graphene surface of GFET in order to avoid unwanted non-specific binding, reduce electrical noise and leakage current. Chemical passivation is crucial to avoid false positive reactions when complex biological analytes are assayed and is of key importance to achieve very low detection limits in the presence of interference buffer solution background with high ionic strength.[163, 164] For example, GFETs exhibit a pH response about 12-45 mV per pH that can be ascribed to uncontrollable and random surface contaminations and/or defects (introduced during device fabrication or storage) that react with protons. However, as shown in Figure 4a and 4b, these defects can be neutralized by covering the surface with fluorobenzene, consequently reducing sensitivity substantially down to <1 mV per pH.[17] A clean GFET therefore acts as a reference electrode that is sensitive to only the change of electrostatic potential in aqueous electrolytes, unless a chemo-adsorption or a physico-adsorption of charged ions is considered.[17] Besides the fluorobenzene, BSA and Tween 20 are commonly used molecules to self-assemble on the graphene surface to rule out possible false nonspecific reactions and thus maximizing biospecific binding.[24, 163]

In fact, to achieve the suppression of certain property or performance of GFET, graphene surface passivation techniques include not only self-assembly, noncovalent bonding, or layer encapsulation, but also chemical reaction or even introduction of defects. For example, the Raman results in Figure 4c shows that the hydrogen radical plasma introduces a low density of H-sp$^3$ defects into the monolayer graphene lattice upon continuous exposure (more than 1 s).[52] Further hydrogenation reduces $\mu$ (and $G_{min}$, not shown) of graphene from ~1250 cm$^2$ V$^{-1}$ s$^{-1}$ down to ~750 cm$^2$ V$^{-1}$ s$^{-1}$ (2-5 s), after which $\mu$ stabilizes at 450 - 660 cm$^2$ V$^{-1}$ s$^{-1}$ (5-30 s). In the meanwhile, $k^0$ first sharply drops from 6.77 × 10$^{-4}$ cm s$^{-1}$ down to ~1.70 × 10$^{-4}$ cm s$^{-1}$ (within 5 s) and then stabilizes at 1.50 × 10$^{-4}$ cm s$^{-1}$ after 30 s of hydrogenation (Figure 4d). Thus, these studies on the interplay between the electrochemical activity and the electrical in-plane transport of graphene, indicate that the addition of one H sp$^3$ defect per 100 000 carbon atoms reduces the electron transfer rate of the graphene basal plane by more than 5 times while preserving its excellent $\mu$ to a large extend.[52] Indeed, quantum capacitance measurements demonstrated that the mild hydrogenation within 1-5 s effectively depresses the average density of state (ADOS) in graphene (Figure 4d). These insights into using hydrogenation to change the electronic structure of graphene, and predict well the electrochemical activity suppression based on the non-adiabatic theory of electron transfer.[52] Such an interesting coordination suggests hydrogenated graphene as a potential approach to improve the sensitivity of GFET (with lower electrochemical current) going beyond previously reported GFET.

## 4. Operation tuning of GFET for biochemical sensing

Unlike traditional electronic devices that are carefully processed to remove any possible surface contaminants or trap states, graphene biosensors have to be functionalized with elaborated biopolymers as receptors against target biomolecules. To achieve *in-situ* biosensing, graphene biosensors are handled under liquid environments, i.e., far from ideal operation conditions of electronic devices, which are normally deliberately sealed against moist in the ambient air. Thus, the sophisticated sensing conditions in aqueous solutions have posted great challenges for achieving reliable operation of electronic devices with optimized performances. That is, biosensors have to overcome the interferences of external noise, as well as the ionic atmosphere screening in the physiological solution to detect a trace of the charges of the biomolecules, which call for interdisciplinary research efforts not only in materials and chemistry/biology, but also in physics and semiconductor devices. In the following, we will highlight our recent progress along these lines based on ambipolar operation near the CNP and high frequency measurements towards low noise and highly sensitive biodetections.

### 4.1. Ambipolar frequency multipliers

By changing gate voltage $V_{GS}$ of GFET, the Fermi energy of graphene sheet (i.e., the electrochemical potential of the charge carriers) can be modulated and the type of charge carriers that flow in the graphene channel can be continuously tuned from holes to electrons, yielding the so-called 'ambipolar behavior'. Therefore, as a peculiar characteristics of GFETs, the ambipolar behavior of graphene stems from its lack of an inherent band gap,[15, 49] and can be utilized to design electronic circuits,[165] such as frequency doublers and/or multipliers with excellent performance.[50, 166, 167]

The first frequency-doubling biosensor device was implemented by biasing an ambipolar GFET in a common configuration (Figure 5a). That is, the input sinusoidal voltage applied to the electrolyte gate with frequency *f* can be amplified and sampled at the drain contact at frequency 2*f* (Figure 5b). The strong electrolyte-gate coupling results in a high output purity (more than 95%) and a high unity gain of the frequency 2*f* sine wave at the drain electrode (Figure 5c).[20] An improved drift characteristics combined with low 1/*f* noise by sampling at doubled the signal frequency, indicates that the graphene frequency doubler is promising for biochemical sensing. Additionally, due to the cleaning effect and the suppression of the resistance drift (either from the graphene channel or from the contact), such GFET operated at frequency-doubling mode shows electrolyte-gate voltage referred input drift less than 0.1 mV h$^{-1}$ during a one-week period. This is equivalent to or exceeds the drift performance of diamond-, high-performance Si-, and conventional graphene-based biochemical sensors. Thus, electrolyte-gated GFET in frequency-doubling mode provides much high flexibility and tunability for realizing biochemical sensors with great reliability and stability.[20]

**4.2. Biosensing near the neutrality point of graphene**

Conventionally, in order to achieve maximum sensing response, graphene transistors are operated at the maximum transconductance point. However, it is found that the electronic noise is unfavorably large, and thus representing a major limitation for realizing the next generation graphene biochemical sensors with ever-demanding sensitivity. Interestingly, for graphene supported on SiO$_2$/Si substrate, the electrical noise exhibits a (local) minimum at the neutrality point with the lowest density of states.[110] Biasing at such a low-noise CNP has been advantageously designed into the graphene Hall bar devices that have demonstrated the steepest sensing response with respect to Hall resistivity.[16, 33] Nevertheless, such sensor devices in Hall geometry require an elaborate magnet setup, which makes it unsuitable for integration and portable application.

The first example of graphene sensors operated near the low-noise neutrality point in a transistor geometry is realized by making use of the unique ambipolar behavior of graphene. That is, applying a sine wave on the gate voltage of GFET near its CNP and monitoring the output current under a constant bias voltage $V_{bias}$ (Figure 5d, e),[24] which exhibits significantly reduced electronic noise, as the current noise PSD is at its (local) minimum. It is worth noting that this electronic noise reduction is achieved without compromising the high GFET sensing response and thus resulting in a significantly increase in SNR, compared to a conventional GFET (Figure 5f). In order to explore the broader range of biochemical sensing applications of the near-neutrality point operated GFET, HIV-related DNA hybridization was selected as the test bed and ultrasensitive detection at picomolar concentrations can be received,[24] with the label-free and portable prospects of graphene nanoelectronics devices.

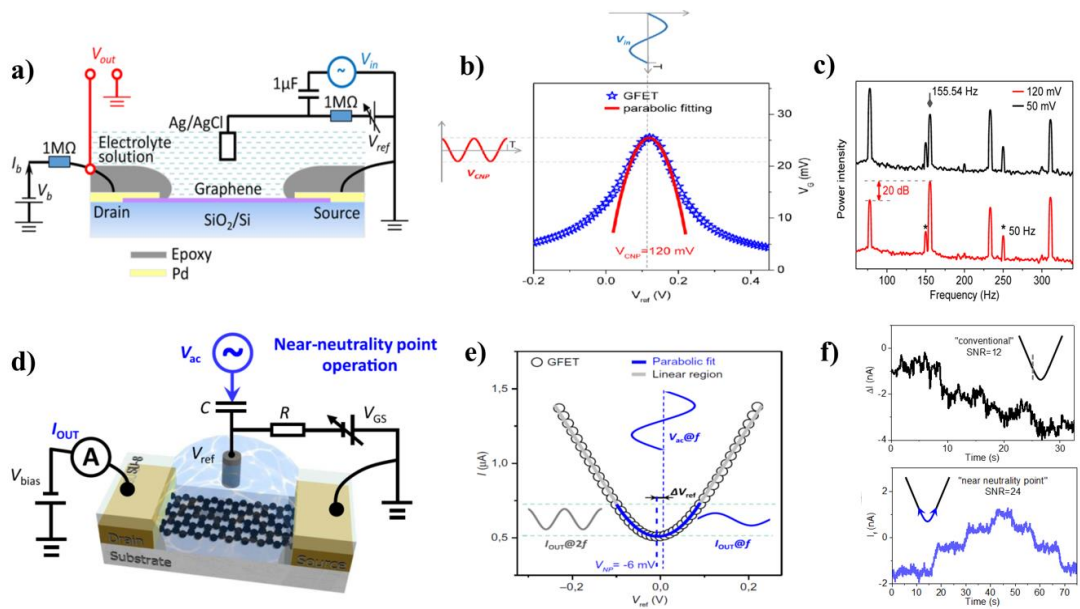

**Figure 5**. (a) Illustration of a liquid-gated GFET operated in the frequency-doubling mode. (b) Transfer curve $V_{GS}(V_{ref})$ (blue stars) and parabolic fitting results (red) of GFET around CNP. (c) GFET's power spectrum at $V_{ref}$= 50 mV (black) and 120 mV (red). (d) Illustration of a liquid-gated GFET operated near CNP. (e) Transfer curve $I(V_{ref})$ (black) and parabolic fitting results (blue) of the GFET near CNP, and the linear fit (gray) away from CNP. (f) Comparison of the SNRs of the GFETs operated in traditional mode (upper panel) and close to CNP (lower panel) with 200 µV step gate voltage variation, respectively.

### 4.3. Overcome the Debye screening effect with RF-operated GFETs

Conventional GFETs are able to sensitively response to the adsorption of biomolecules with charge close to the surface of graphene. However, as mentioned in Section 2.3.4, GFET-based biosensors (and ISFETs in general) have the problem of ionic screening due to the mobile ions in the solution, also known as the Debye screening effect.[119, 120] At physiological conditions with a Debye length around 0.7 nm, the charges on the target molecules are heavily screened. It is difficult or even impossible to detect if the distance between the graphene surface and the charged biomolecules exceeds several Debye lengths (Figure 6a). For better performance, various approaches, especially high-frequency measurements have been reported to circumvent the Debye screening effect as previously discussed in Section 2.3.4.

As a promising alternative strategy to overcome the Debye screening effect in physiological conditions, measuring at high frequencies is able to achieve improved sensitivity while no special design or engineering of the sensor environments or the receptor molecules is needed.[71] Indeed, graphene is potentially suited for high-frequency applications owing to its exceptional high carrier mobility.[22] For example, the intrinsic cut-off frequency ($f_T$, the highest frequency of a FET under RF) of GFETs is 100-300 GHz,[168-170] which surpasses the best silicon based FETs.[171, 172] In order to enlarge our understanding on the RF properties of GFET, in particular sensing in liquid, an electrolyte-gated GFET operated at ≈2-4 GHz has been demonstrated (Figure 6c).[71] The sensitivity to the load is optimized by utilizing a tunable stub-matching circuit implemented on printed circuit boards (PCB) with ground planes and coplanar waveguides (CPWs) (Figure 6c). Using reflectometry technique, the reflection coefficient $S_{11}$ under a range of electrolyte-gate voltages can be achieved and analyzed. According to the kinetic inductance and negligible skin effect,[110] atomically thin large area graphene behaves as a wideband resistor. However, at RF the device resistance cannot be directly measured because of the large shunt capacitance in conventional RF GFET, which has a significant influence on the RF performance and hinders the extraction of the intrinsic parameters of graphene.[71] Due to the special frequency dependent properties of the electrolyte, the properties of GFETs can be tuned by means of liquid gating without a significant spreading of the RF signal. Therefore the gate dependent resistivity of graphene at RF can be extracted by considering an *RC* dissipative transmission line mode, which perfectly matches its DC counterparts in the full range of gate voltage sweeping.[71] Due to its wide bandwidth (100 MHz) and a significant reduction in 1/*f* noise at RF, such RF GFET achieved ultrafast measurements (10 ns time resolution in the electrolytes) with good detection limits (Figure 6d).[71] As a proof-of-concept for ultrafast sensing in liquid environment, this work initiates the further study on a new generation of biosensors in the field of environmental, biomedical, in particular with great potential to applications from POC medical diagnosis to neuronal sensing.[89-91, 173] In addition, AC field also has great effect on the liquid and leads to micro-nanoelectrokinetic phenomena e.g., dielectrophoresis and electroosmosis,[174] which could be used to classify, manipulate, and concentrate different biomolecules and nanoparticles at the strongest field to further improve the detection limit.

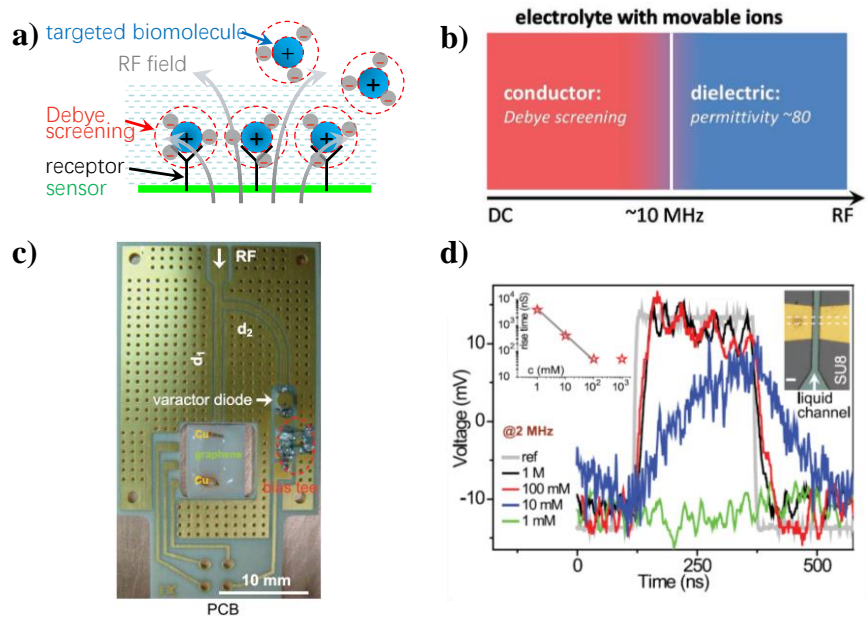

**Figure 6**. (a) Debye screening effect prevents the charged biomolecules from delivering an appreciate sensing signal outside the Debye length. (b) Debye screening effect mainly occurs under DC and low frequency (<10 MHz). In contrast, the electrolytes behave as dielectrics under higher frequencies. (c) Optical image of a PCB used for the impedance match between load impedance (typical resistances of 1-100 kΩ) and the transmission line impedance at 50 Ω. (d) GFET responses to KCl solutions with different concentrations under a liquid gate of 500 ns square signal with 200 mV voltage. Right inset is an optical image of the SU-8 sealed GFET. Scale bar: 10 μm. The left insert indicates a linear $1/C$ behavior, i.e., the measured rise times as a function of the KCl concentration $C$.

### 4.4. Other electronic tuning approaches

Due to the peculiar electronic band structure with linear dispersion,[15, 49] graphene lacks a band gap and is of metallic nature. Thus, GFET devices cannot be turned off at room temperature. For digital applications where an ultra-low power consumption is required at the off state ($I_{off}$) of FET, chemical treatment, nanomesh, nanoribbon, and nanoring,[75] have been widely applied and proven as rational designs to open a bandgap in graphene for achieving an improved $I_{on}/I_{off}$ ratio. However, bandgap engineering is more crucial for large-scale integration of sensor devices, where power consumption and heating dissipation due to off-state current are of key importance. For general applications, rather, the performance of sensor devices depends primarily on sensitivity and selectivity. To meet the ever-demanding requirements on sensitivity, nanopores and nanometer-sized gaps based on translocation blockage current and tuning current, respectively, have been proposed and realized for ultimate single molecular detection. GFET-based DNA sensors in the form of a nanopore in the center of a graphene nanoribbon FET have also been fabricated.[175-177] Translocation of the ssDNA molecules through the graphene nanopore results in variation in the conductance of the GFET.[178] Another commonly utilized technology for single molecule study uses graphene break junctions. Compared to the most common break junctions, which consist mainly of gold as electrode material, monolayer graphene grants an easy access for not only optical and scanning probe imaging, but also *ex-situ* gating experiments owing to its ultimate thickness, flexibility and robustness.[179]

**5. Applications of graphene biochemical sensors**

Graphene nanoelectronic devices provide a versatile platform for a wide range of biosensing applications.[180] Particularly, the design and fabrication of the first GFET[15] has inspired considerable theoretical and experimental studies on the applications of GFET for high-performance and label-free biochemical sensing[16, 17] on the presence, adsorption, and reactivity of gases and ions, DNA, proteins, cells and tissues. In this Section, we review the development (especially our recent achievements) on GFET-based sensors in meeting the social/scientific needs on biochemical sensing for environmental monitoring and food safety, human health and medical diagnosis and life science research.

**5.1. Graphene biochemical sensing for environmental monitoring and food safety**

The growth of population puts an increased requirement for high quality living conditions, and environmental monitoring and food safety appear to be serious problems in front of our society. As a new emerging material with unique properties, graphene shows its potential as highly sensitive and biocompatible material for gas and ion sensors that could be used for food safety and environmental monitoring. Nonetheless, it is now widely accepted that the previous reported sensitive responses of graphene to the presence of gas molecules or ions could be due to the sensitivities of polymer contaminations or defects introduced during graphene device fabrication and/or storage, and clean graphene should be insensitive. Indeed, after removing the possible surface contaminations by annealing at high temperature of 400 °C in Ar/H$_2$ atmosphere,[181] the cleaned graphene surface is insensitive even upon the exposure to 1000 ppm NH$_3$[34] or to 100 ppm dimethylmethylphosphonate (DMMP) vapor.[40] We have also discussed in Figure 4a and b that the passivation of graphene transistors using fluorobenzene molecules, results in an inert sensing response to the change of pH values in the buffer solutions. Only via deliberated surface functionalization of graphene, highly sensitive and selective detection of targeted biochemical molecules can be achieved.

**5.1.1. Ethene gas sensors**

Monitoring the concentration of ethene is critically important for the storage and transport of crops to avoid ethene-induced spoilage, i.e., when the concentration of ethane rises, the resulting deleterious effects will lead to over-ripeness or even spoilage of crops.[182] Particularly, the highly diffusive and relatively unreactive ethene induces deleterious effects at very low concentrations of parts per billion (<100 ppb). In this respect, the development of ultrasensitive detector systems with good ethane selectivity for *in-situ* monitoring of the ripening processes of crops is highly desired. Copper(I) compounds are able to selectively detect ethene either optically by combining the polymers with fluoresce[183] or electrically by using carbon nanotube network as conductive channel.[184] However, the good selectivity, reasonable sensitivity (down to 500 ppb) are offset by the poor reproducibility of the sensors due to the use of the randomly placed single-walled carbon nanotubes (SWCNTs) and the inhomogeneous crystallites of the complex distributed among the conductive nanotube network.

The use of 2D graphene materials instead of nanotube networks allows for the exploitation of the ultrahigh sensitivity granted by the carbon allotropes with all-surface-atom makeup like graphene and SWCNTs, but without the aforementioned practical shortcomings. Indeed, GFET functionalized with copper complexes (Figure7a) is able to detect ethene at a concentration of low part per billion (ppb, Figure 7b).[19] In order to understand the chemical interactions between molecules, which leads to the sensing response, a systematically engineered series of copper complexes with deliberated varied dipole moment has been designed.[185] GFET is adopted to harvest the molecular dipole fluctuations when the copper complexes undergo a chemical reaction upon the introduction of ethene. In this respect, GFET is a promising platform for studying the interplays between molecules. Remarkably, it is possible to track the chemical reaction and probe into the mechanism that was, until now, out of reach. In Figure 7c, by using a Langmuir adsorption isotherm, the equilibrium constant $K_D$ can be extracted, which is useful for deriving a plausible reaction mechanism. With further attention on the sensitivity and/or reproducibility, these small GFETs have the potential to be widely applied in the greenhouses as well as in the storage and transportation of crops to meet the demand for a safe and stable supply.

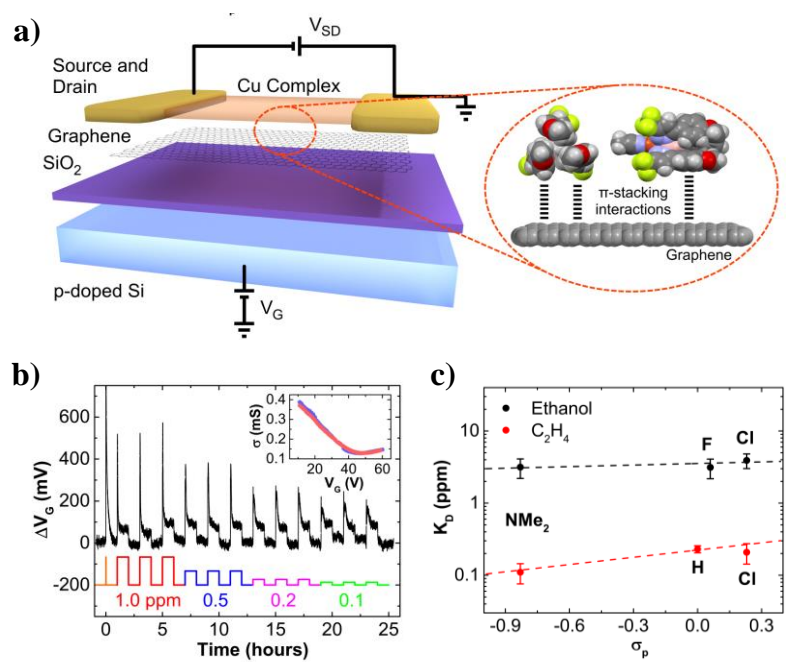

**Figure 7**. (a) Schematic representation of a GFET device functionalized with copper complex. Zoom: The space-filling projections of the acetonitrile adduct of the copper complex stacks on graphene. (b) GFET response to ethene with different concentrations. Inset: The back-gate voltage ($V_G$) versus graphene conductance ($\sigma$) on SiO$_2$/Si substrate before (black) and after (red) the functionalization of copper complex. (c) The equilibrium dissociation constant $K_D$ of ethylene (red) and ethanol (black) on GFET as a function of the Hammett parameter $\sigma_P$ of the substituent on the sensitizer ligand.

### 5.1.2. Ion sensors

Ion sensors based on highly sensitive GFETs have potential applications especially in medical diagnosis and food industry that require glass-free ion measurement with small size, high performance, and/or flexibility.[186] After systematically studying the response of GFET to a large range of pH solutions, it is clear that graphene is intrinsically insensitive to pH.[187] Whereas an appreciable pH response (~40-50 mV per pH) is expected if the graphene surface is covered with an ideal $Al_2O_3$ layer, which can be protonized and deprotonized through the terminal hydroxyl groups, yielding a layer with charge density related to the proton concentration in solution.[188] Similarly, by anchoring a crown ether (dibenzo-18-crown-6-ether) with high affinity to $K^+$ on the surface of graphene via π-π stacking, a desirable sensing response can be recorded when increasing ion concentrations in a wide range from 100 μM to 1 M. Whereas as-fabricated GFETs exhibit only a weak sensitivity (about 3 mV per pK).[18]

Due to the hazardous effect of heavy metals (e.g., Cr, Hg, Pb, Cd) on environment and health, highly sensitive and selective heavy metal sensors have attracted a wide research interests.[189, 190] For example, graphene surface can be functionalized with a self-assembled 1-octadecanethiol monolayer for the detection of $Hg^{2+}$ at 10 ppm,[191] the sensitivity of which can be attributed to the firm binding between the $Hg^{2+}$ and the thiol groups of the 1-octadecanethiol. Moreover, prototype devices decorated with DNAzyme aptamer are capable of detecting $Pb^{2+}$ down to 37.5 ng $L^{-1}$ in real blood samples.[192] Besides the basal plane, the edge of graphene and the underneath substrates are also of key importance on determining its sensing properties (as well as the electronic, chemical, and physical properties). For example, defective and holey rGO might contain abundant edge defects due to enzymatic oxidation and hydrazine reduction processes, resulting in a selective and sensitive electronic detection of hydrogen, particularly if functionalized by using Pt nanoparticles.[147]

**5.2. Graphene biochemical sensing for human health and medical diagnosis**

Graphene electronic biosensors for POC applications may have a significant societal impact for medical diagnosis, including DNA and protein biomarker detection. The engineering of graphene-protein interfaces is crucial for efficient sensing.[193] For example, by bounding enzymes (e.g., glucose oxidase) onto the graphene surface and integration into a microfluidic device, a graphene-based POC biosensor platform for glucose detection of diabetes patients is suitable for home use.[194] A rGO FET with functionalization of PSA monoclonal antibody was reported able to detect a complex biomarker (i.e., prostate specific antigen/α1-antichymotrypsin) in prostate cancer diagnosis in femtomolars concentration.[45] A rapid POC sensor based on a dielectric-gated and resonance-frequency modulated GFET was able to detect the Ebola glycoprotein with a sensitivity of ∼36-160% and ∼17-40% for 0.001-3.401 mg L$^{-1}$ at high and low inflection resonance frequencies, respectively.[46] By using bioactive hydrogels as the gate material and encapsulating biospecific receptors inside, enzymatic reaction can be effectively catalysed in the confined microenvironment, enabling real time, label-free detection of biomolecules (e.g., penicillin down to 0.2 mM). Bioactive hydrogels are able to significantly reduce the nonspecific binding of nontarget molecules to graphene channels as well as preserve the activity of encapsulated enzyme for more than one week, which is important for POC application.[42] ssDNA strands can be detected through hybridization with complementary ssDNA anchored on the GFET surface.[195] Such GFET DNA sensors are able to distinct the hybridization of DNA with single-base precision[196] or distinguish the four DNA nucleobases based on different dipole field upon their adsorption.[152] Multiplexed CVD grown GFET-based DNA sensor arrays can also be manufactured and acted as an electrophoretic electrode not only for immobilization of site-specific DNA but also for detection of complementary DNA with concentration of 100 fM.[61] Aptamers with high specificity and affinity to certain biomolecules, are another pre-selected analytes for novel GFET sensors.[197-199] Such graphene-aptamer complexes have been successfully used to detect immunoglobulin E (IgE) proteins,[197] Hg$^{2+}$,[200] small molecule steroid hormones,[201] interferon-gamma (IFN-$\gamma$),[202] and adenosine triphosphate (ATP).[41]

A recently developed novel operational scheme of biosensing near the neutrality point of graphene, is able to further optimize the sensing performance as a result of its extremely low noise level and excellent detection limit, compared to traditional conductance measurements. The simple sensing scheme is achieved by operating a GFET in an ambipolar mode close to its CNP, where the low-frequency 1/$f$ noise is found to be minimized (see also Section 4.2). Using specifically designed aptamers anchored on the surface of GFETs (Figure 8a, b), the GFET operated in ambipolar mode is capable of detecting an HIV-related DNA hybridization process at picomolar concentrations.[24] The graphene surface was first functionalized with pPNA aptamer that can hybrid with the target complementary HIV ssDNA (see also Section 3.1.2) and passivated with self-assembled Tween 20 to rule out possible false non-specific positives (Figure 8c). When operated near its neutrality point, GFET functionalized with the pPNA are able to detect 11mer ssDNA at a limitation of ~2 pM 1 mM PBS with an RMS SNR of 1 (Figure 8d). The same HIV related ssDNA can also detected by using GFET operated in frequency-doubling mode (see also Section 4.1).[20] It is expected that sub-pM sensitivity can be achieved if carefully controlling the Debye screening. Therefore, GFETs operated near the neutrality point or in frequency-doubling mode are able to promote the application of low-noise graphene sensors for biomarker detections, which are at the core of biochemical sensing for human health and medical diagnosis. In addition, biomolecules (e.g., short RNA) may undergo conformational change (e.g., melt or unfold) when attached on graphene surface, which could complicate design but also imply possibility of manipulating the properties of surface-bound biomolecules.[153]

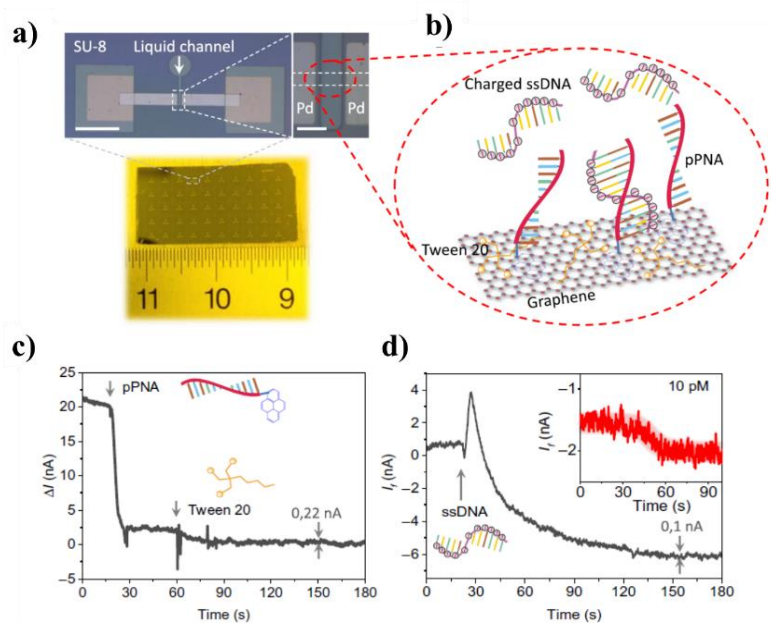

**Figure 8**. (a) Optical image of a GFET array and an individual GFET device. (b) Schematic representation of the binding if a negatively charged complementary ssDNA molecule to a pDNA molecule that is noncovalently bounded on the surface of graphene. The self-assembly of Tween 20 on the graphene surface is applied to avoid the non-specific adsorption on the sensor surface. (c) Variations in the sensor $\Delta I$ during traditional GFET measurement upon the self-assembly of pDNA and Tween 20. The RMS current noise is estimated to be 0.22 nA. (d) Variations in the sensor $I_f$ operated around CNP upon introducing 1 nM and 10 pM (inset) complementary ssDNA show a current noise of 0.1 nA.

## 5.3. Graphene biochemical sensing for life science research

New generation of neuroprosthetic sensor development requires advances in material science, solid-state sensors, and actuators to further improve signal detection capabilities with good stability in biological environments and compatibility with living tissue. The drawbacks associated with the conventional silicon technology[203], such as its mechanical mismatch,[204] instability in liquid environments[205] as well as the high electrical noise,[206] triggered the research on alternative technologies and materials.[23, 35, 36] Besides the superior FET performance compared to the most semiconductors due to its excellent electrical properties,[49] graphene also possesses good chemical stability[207] and biocompatibility,[208] which is beneficiary for both integration with biological systems and operation of GFETs without dielectric protection. In addition, graphene devices integrated with flexible substrates opens up the possibility of developing flexible and soft devices, a crucial requirement for reducing tissue scarring and damage from implantation.[209, 210] The first realization of using GFETs to detect electrogenic signals of cardiomyocyte cells (Figure 9a) was achieved with a SNR >4.[211] Further development of the array of GFETs (Figure 9b) towards cellular electrophysiology comes together with the advances in large-scale CVD growth technology. In such GFETs, graphene as a conductive channel can not only detect the presence and the activity of the cells, but also act as conductive electrode to transduce stimuli into the cells.[195] In general, GFETs are regarded as active components with tunable and functional performance in comparison to the passive MEAs technology.[212] Moreover, the size of graphene devices can be more aggressively scaled down while preserve its super electronic performance if only keeping the *W/L* ratio.[213]

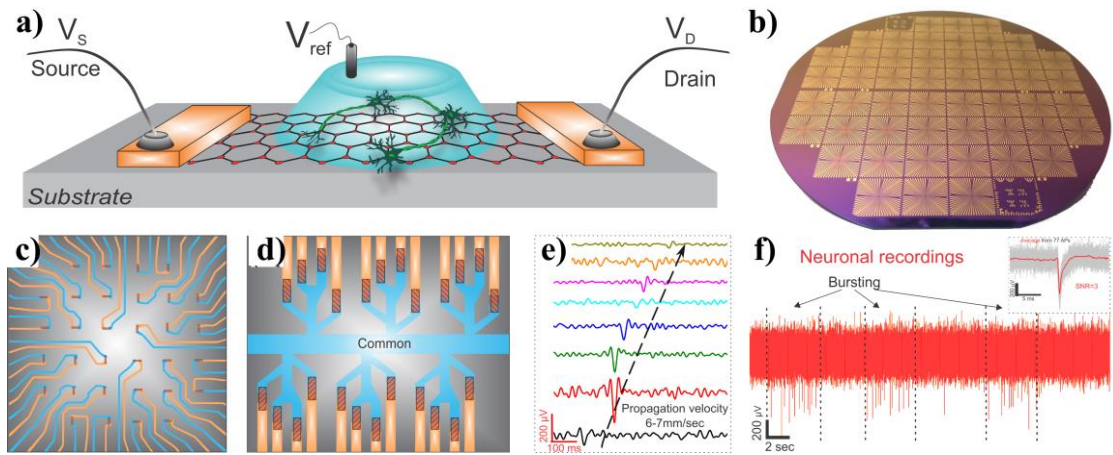

**Figure 9**. (a) Schematic presentation of using a GFET for the detection of single or multiple electrogenic cells. (b) Optical image of a wafer has numerous chips with arrays of GFETs. (c) An example of a GFET array where each transistor has a separate source and a separate drain connection, while (d) shows an example when a single common source is used to contact multiple transistors per chip. (e) 8 time traces of extracellular activity of cardiomyocyte cells with visible time delay between the transistors that can be consequently recalculated into signal propagation velocity. (f) Neuronal recording time tracking features of a burst of intrinsic neuron, when neurons show alternative high frequencies (bursts) and low frequencies, intermittent, spikes. The inset shows an average AP (red) of 77 individual APs (gray).

For the detection of the action potentials of cells cultured on graphene surface, mainly two designs of GFET arrays have been developed and reported. One design uses two contacts for a single transistor (Figure 9c). The another design implements a single common electrode that works as a common source or drain for all transistors at the same time.[35] Both designs have their advantages and drawbacks. In the first approach it is possible to specifically tune the operational parameters for each transistors, while in the latter case all transistors are configured with the same operational parameters. Advantage of the common electrode design, however is the possibility to measure a larger number of FETs per chip to study the cellular signal propagation in better resolution. Typically, the action potentials propagation across the electrically active cells can be detected by measuring the flow of current through each GFET in the array. The cell-graphene interface variations and the action potential propagation across the beating cellar network can be depicted as spikes in current trace. The action potentials of electrically active cardiomyocyte cells could be detected and monitored across GFET arrays (see Figure 9e), making it possible to calculate signal propagation velocity as well as to track the spatial signal propagation. Even in the early stage of its development, the detected signals together with the related transistor noise exhibit a SNR better than ten, which surpasses that of the state-of-the-art techniques based on planar FET, MEAs, and nanowire FET.[211, 214-216] One challenge for modern bioelectronics is to record and stimulate the neurons' extracellular or even intracellular potentials with branched transistors.[217] The first *in vitro* neuronal signals (the action potentials, APs) with clear bursting detected by GFETs is shown in Figure 9f,[36] although the APs of extracellular neurons are small compared to those of heart tissue[218]. Based on the transconductance of transistor, spikes of gate voltage can be deduced from the spikes of current, yielding a value of 900 μV with a RMS of 50 μV.[35] For future prospective, integrating high-performance GFETs with flexible substrates may initiate a breakthrough on bioelectronics, especially for electrically neural prostheses.[219]

Interestingly, one of the advantages of the emerging 2D materials beyond graphene is their ultimate thinness, allowing them to be integrated into extremely thin and even soft shells, creating a promise for a fully 2D based neuroelectronic implantation. One of such materials is a 2D $Ti_3C_2$-MXene that is fabricated by selective etching of Al in the $Ti_3AlC_2$ three-dimensional structure. It was found experimentally that the material is sensitive to neurotransmitters, such as dopamine,[220] and can serve as a conductive microelectrode,[221] thus representing an intriguing opportunity for building future bio- and neuroelectronic interfaces.

## 6. Challenges, perspectives and conclusions

### 6.1. Reliability and reproducibility

Improving further sensitivity with ever-demanding reproducibility and reliability should be the focus of future direction for GFET-based biosensor.[18, 20] Although graphene-based electronic devices with superior performances have been achieved, the reproducibility and reliability of GFET biosensors were not always studied or achieved, which represent a big challenge for the development of next generation GFET sensor devices.

To date, CVD graphene grown on Cu with meter-length crystals has been achieved in laboratory,[108] which opens the window towards industrial production of high-quality graphene with mobilities over $10^4$ cm$^2$ V$^{−1}$ s$^{−1}$. In order to scale up the fabrication of single-device into wafer-scale, mass producing of large-scale graphene with a well-defined atomic structure, including disorders, defects, impurities, heteroatom and adatoms, is highly desired. Along this direction, novel approaches to minimize the flaws or fluctuations of epitaxially grown graphene[222, 223] and CVD grown graphene have been actively pursued.[224-226] On the other hand, transfer of graphene onto device-compatible substrates is an indispensable fabrication step, and represents another critical challenge. The introduction of defects during the transfer process, results in a low yield of the graphene devices and calls for the exploration of effective, large-scale transfer approach.[227, 228] Indeed, conventionally used polymers for transferring 2D materials such as PMMA – tends to attach on the surface of graphene irreversibly, leading to various unwanted chemical contaminations.[229-231] Therefore, the influences of these possible polymer residues is necessary to be taken into account on the performance of sensor device as they impede the graphene surface functionalization. In this respect, decent polymer-free transfer methods are highly demanded,[232-234] such as using a biphasic oil-water interface for clean transfer.[235] In the meantime, the transfer of crack- and fold-free large area graphene sheet is still a tricky skill, although combined with nano/microfabrication technique, high-throughput transfer of graphene and large scale fabrication of GFET arrays (52 devices per 4-inch wafer) was achieved for more reproducible performance of the GFETs.[236, 237] Direct growth technology on arbitrary substrates[238] is an alternative way to avoid this issue caused by graphene transfer, but generally resulting in low quality of graphene compared to metal catalysts. On-surface bottom-up approach is promising for achieving atomically defined GNRs, offering additional opportunity to control the microstructure of graphene.[238]

Conventionally, SEM, AFM, STM and Raman spectroscopy, are wildly adapted in laboratory as crucial tools to identify the structure and the physical property of graphene. However, such means are either invasive or not applicable to the characterization of the electrical properties of large-batch graphene films. In this respect, novel techniques with non-destructive, high accuracy and speed, are urgently needed. In pursuit of rapid property evaluations (conductivity, uniformity, continuity, etc.) of large-area graphene, researchers have developed terahertz time-domain spectroscopy[239] and microwave resonator[240] for effective characterization of graphene conductivity (and even quantum capacitance) without physical contact. The development and evolution of such characterization technologies highlight great opportunities in both scientific research and business.

**6.2. Perspectives and conclusions**

Applying graphene-based electronic devices for biochemical sensing applications, including environmental monitors, portable POC devices for remote diagnostics, and even for DNA sequencing technologies, etc. has risen a vast interest from scientific community, industry, and society.[89-91] Although in principal and also experimentally demonstrated GFETs are able to reach ultimate single molecule sensitivity and various prototype forms of GFET chips were developed,[89-91] the research outcomes have not reach the market yet.[241] Smart GFET biochemical sensors will be an impressive prospective, those can be wearable and wireless [29, 242, 243] with low energy consumption and low maintenance cost for event-based, real-time monitoring in pervasive healthcare IoT applications.[173]

In an attempt to compete with current mature material in the market, GFET sensor devices have to stand out in both cost and performance. Compared to exfoliated graphene, high-quality CVD-grown graphene is promising for large scale production of GFETs. Nevertheless, the consumption of the substrate and energy during high-temperature CVD synthesis are not cost-efficient. To accelerate the commercialization of high-performance GFET devices, reuse of metal by optimizing transfer technique and cold-wall CVD (e.g. PECVD at temperatures below 500°C,[244] would be beneficial to promote the mass-production of graphene at industrial level.

Up to now, large-scale, high quality graphene sensors with average mobility ~5000 $cm^2$ $V^{-1}$ $s^{-1}$ can be routinely fabricated.[44, 245] Nevertheless, the reported electronic characterizations of GFET biochemical sensors are still behind expectation (see also Section 2.3). To improve the electrical performance of GFETs towards their theoretical predictions, basic outlines include doping, surface treatments, edge contacts, modifications of substrates, graphene, and the interaction between them. For instance, it is well-known that on standard oxide substrates, such as $SiO_2$ or $Al_2O_3$, graphene devices are highly disordered and exhibit inferior characteristics. In this regard, h-BN with atomically flat surface free of dangling bonds, is an ideal substrate to match and suppress disorder in graphene electronics to achieve theoretical performance.[85]

In summary, this article features recent progresses on research efforts devoted to understanding the sensing mechanisms of GFETs, to functionalizing the surface of graphene with recognition groups to unblock its selectivity and sensitivity towards targeted molecules, and to conditioning the sensor devices under optimized operational conditions by utilizing the unique electronic properties of graphene. We firmly believe that graphene holds great promise to meet the high requirements on next generation biosensor development, especially combining the tuning strategies enabled by graphene surface functionalization, multifrequency ambipolar detection and high-frequency operation.

**Conflict of Interest:** The authors declare no conflict interest.

**Acknowledgements**

The authors acknowledge financial support from the National Natural Science Foundation of China, the Swiss National Science Foundation, European Commission Horizon 2020- Research and Innovation Framework Programme (Marie Sklodowska-Curie actions Individual Fellowship No.749671).